\documentclass{article}
\usepackage{spconf,amsmath,graphicx}
\usepackage{times}
\usepackage{url}

\title{Exact Prosody Cloning in Zero-Shot Multispeaker Text-to-Speech}

\name{Florian Lux, Julia Koch, Ngoc Thang Vu}
\address{University of Stuttgart, Institute for Natural Language Processing, Germany}

\copyrightnotice{978-1-6654-7189-3/22/\$31.00~\copyright2023 IEEE}

\begin{document}

\maketitle

\begin{abstract}
The cloning of a speaker's voice using an untranscribed reference sample is one of the great advances of modern neural text-to-speech (TTS) methods. Approaches for mimicking the prosody of a transcribed reference audio have also been proposed recently. In this work, we bring these two tasks together for the first time through utterance level normalization in conjunction with an utterance level speaker embedding. We further introduce a lightweight aligner for extracting fine-grained prosodic features, that can be finetuned on individual samples within seconds. We show that it is possible to clone the voice of a speaker as well as the prosody of a spoken reference independently without any degradation in quality and high similarity to both original voice and prosody, as our objective evaluation and human study show. All of our code and trained models are available, alongside static and interactive demos. 
\end{abstract}

\section{Introduction}

Customizing TTS to speak in different voices with only a short and untranscribed sample is desireable for any spoken human computer interface. While this field remains not completely solved, impressive results were already achieved with early methods in neural end-to-end TTS \cite{jia2018transfer, arik2018neural}. Exchanging a voice however also usually leads to a change in speaking style \cite{wang2018style}. While adopting the speaking style of a reference is also desireable in some scenarios, there are some cases where we want to be able to customize the voice, but retain a very specific speaking style. One particular application of such a technology could be the anonymization of a speaker by changing their voice. Since their prosody however also conveys meaning, there would be information loss, if the prosody is not carried over to the new voice. Another example could be an audio book, where a professional speaker deliberately chooses to speak with a certain prosody. If we would want to customize the reader's voice, or perhaps turn the audio book into an audio play by assigning each character in an audiobook a unique and customizable voice, we would loose the professional voiceacting that has gone into the original. And further, this technique can prove useful in scientific experiments in areas such as psychology or literary studies whenever it is necessary to factor out the influence of different speakers' voices. We go into more detail on such usecases in Section \ref{sec:applications}.

In this work we propose an approach to prosody cloning that works across any voice in a zero-shot multispeaker setting, as long as a reference audio with matching transcript is available. We extract durations, pitch and energy values for every phone from a reference and then overwrite the respective predictions for these properties that a FastSpeech 2 model makes. A similar control mechanism has been proposed in FastPitch \cite{lancucki2021fastpitch}. We contribute to this 1) a robust way of extracting those properties with a lightweight aligner that can be adapted to individual samples and 2) a way of normalizing pitch and energy that allows for the overwriting procedure to be compatible with a zero-shot multispeaker setting by regaining the value ranges for each speaker through an utterance level speaker embedding. This enables us to apply any prosody from any reference with matching transcript to any voice. 
We conduct experiments on the effectiveness of the prosody cloning, its effect on the quality of the resulting synthetic speech, and its impact on zero-shot multispeaker TTS. Using objective metrics as well as human evaluation we show that our method is capable of accurately cloning fine prosodic details even in extreme cases without degrading TTS quality. Furthermore we show that the same prosodic patterns can be applied to any voice, even unseen ones. Finally, we discuss various examples of how our approach benefits real-life applications.
All of our code and models\footnote{\url{https://github.com/DigitalPhonetics/IMS-Toucan}} as well as a demo with pre-generated audios and an interactive demo\footnote{\url{https://toucanprosodycloningdemo.github.io/}} are available.

\section{Related Work}
Approaches on cloning the voice of a speaker have been proposed before, as well as attempts to copy the prosody from a reference to a target utterance. 
However, to the best of our knowledge, a combination of these tasks by cloning prosody from a reference to even unseen speakers has not been attempted before. 
\subsection{Voice Cloning}

To deal with the challenge of zero-shot multispeaker TTS, \cite{arik2018neural} first proposed using embeddings produced by an external speaker encoder as conditioning signal instead of adapting to new speakers by finetuning the TTS model. With this approach, the same speaker encoder model can be used for all unseen speakers without any finetuning and new speaker embeddings can be generated from only a few seconds of untranscribed reference speech from the target speaker. \cite{jia2018transfer} add to this idea by training the speaker encoder on a speaker verification task. \cite{cooper2020zero} notice there is still a quality gap between seen and unseen speakers which they attempt to bridge by using more informative embeddings than the X-vector \cite{xvect} commonly used in speaker verification. Further attempts on closing this gap have been made by \cite{wang2018style, choi2020attentron} who use attentive speaker embeddings to encode the speaking style of a target speaker - which is however contradictory to our intent to completely disentangle speaker identity from prosody - or by \cite{casanova2021sc} using generative flows in decoding. Nevertheless, this task remains not fully solved and the need for high quality multispeaker data is still high.


\subsection{Prosody Cloning}
There is also a fair amount of research on the replication of a speaker's prosody. This includes methods that operate on a global level to achieve a general style \cite{shechtman2019sequence, raitio2020controllable}. A special case is the Global Style Token (GST) \cite{wang2018style}. They use a reference encoder to encode prosody of a reference audio into a fixed-length prosody embedding. While they are able to transfer style of a reference to variable-length target utterances, this method is not suitable for fine-grained prosody modeling on the phone level. Similar to GST, \cite{skerry2018towards} also use a prosody encoder, but feed the prosody embedding directly to the decoder instead of through an attention module. This allows for prosody transfer with more fine time detail than in \cite{wang2018style}. Encoding prosody in a fixed-length vector however still limits controllability to utterance level. There are also approaches to cloning prosody on a fine grained level. This is achieved in \cite{klimkov2019fine}, however only with a single speaker. In \cite{mohan2021ctrl}, explicit prosodic features are extracted from a reference without encoding them into latent vectors, which is similar to our approach. They concatenate extracted values for duration, pitch and energy with encoder outputs and in \cite{torresquintero21adept}, this technique is used during inference for prosody cloning. Their work differs from ours such that they use an autoregressive model which models prosody less explicitly than our FastSpeech 2 \cite{ren2019fastspeech} based model that predicts duration, pitch and energy in dedicated submodules. Further, their approach is constrained to speakers it has seen during training. 

\section{Proposed Method}

  \subsection{System Architecture}
  Since we want to explicitly control pitch, energy and duration, we choose FastSpeech 2 \cite{ren2020fastspeech} as the TTS component in our cloning procedure. Since FastSpeech 2 however predicts pitch and energy values on a spectrogram frame scale, we take inspiration from FastPitch \cite{lancucki2021fastpitch} and use the average pitch and energy per phone as the output for the pitch and energy predictors. They are then upsampled according to the phone durations in order to be decoded. Furthermore we use the Conformer architecture \cite{gulati2020conformer} in both encoder and decoder. We base our implementation on the IMS Toucan toolkit \cite{lux2021toucan} which is in turn based on the ESPnet toolkit \cite{hayashi2020espnet, hayashi2021espnet2}. 
  To condition the TTS on a speaker reference that can be supplied at inference time, we concatenate the output of two different speaker embedding models and add them to the encoder output, as suggested by \cite{anonymization}. One follows the ECAPA-TDNN architecture \cite{ecapa}  and the other uses the X-Vector architecture \cite{xvect}. Both are trained on Voxceleb 1 and 2 \cite{Nagrani19, Nagrani17, Chung18b} using the SpeechBrain toolkit \cite{speechbrain}. To convert the spectrograms we get from the synthesis into waveforms, we use the HiFi-GAN architecture \cite{kong2020hifi}. To get the pitch curves for training and inference, we use an open-source API\footnote{\url{https://github.com/YannickJadoul/Parselmouth}} to Praat \cite{boersma2001praat}.

\subsection{Cloning Mechanism}
\begin{figure}[ht]
    \centering
    \includegraphics[width=\linewidth]{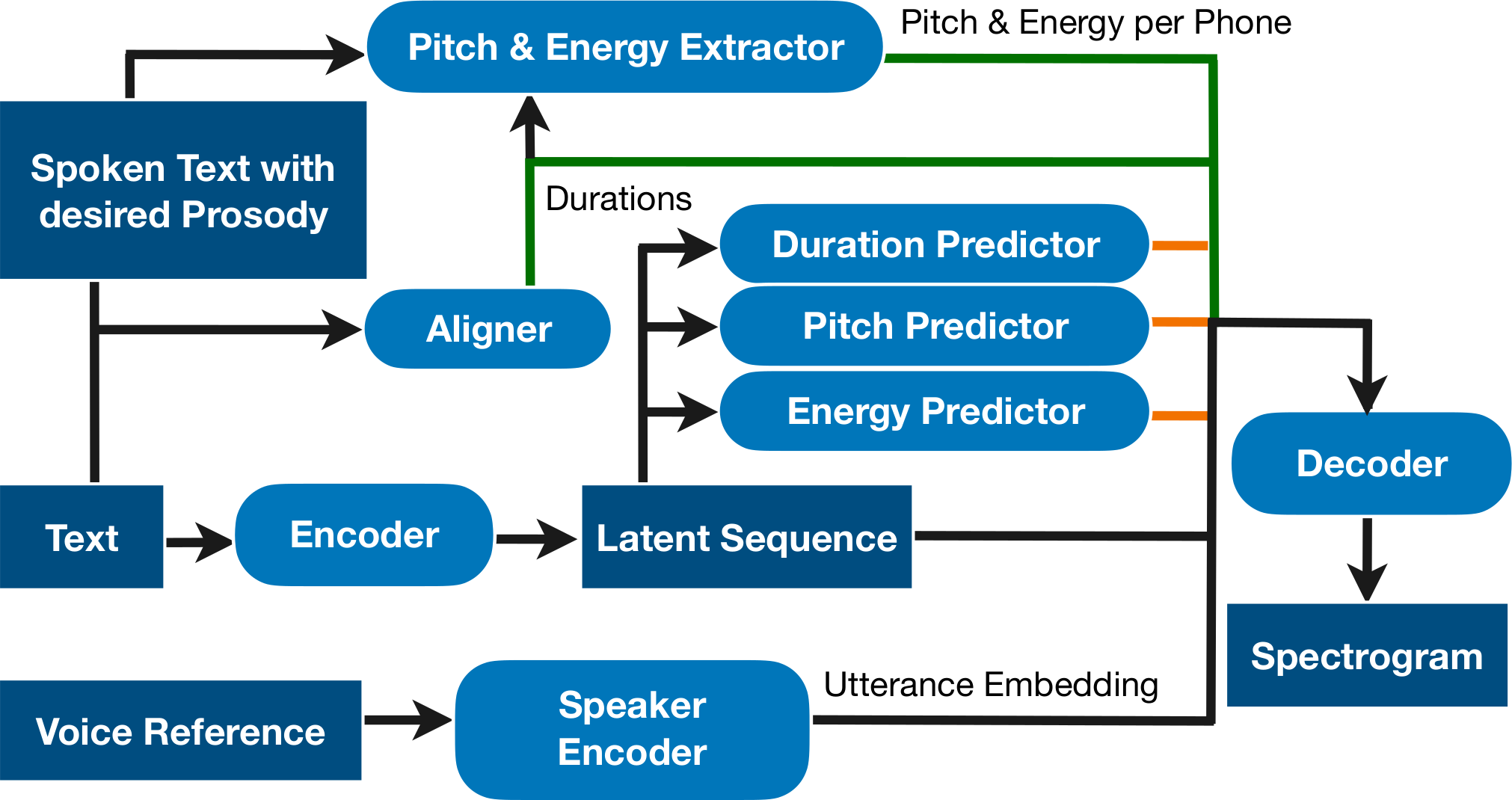}
    \caption{Schematic overview of our proposed cloning procedure. The orange lines are overwritten by the green lines. Prosody reference and voice reference can be the same, but they don't need to be. The utterance embedding sets the normalized pitch and energy values into context with regards to the speaker.}
    \label{fig:overview}
\end{figure}

Since we predict durations explicitly and one value for pitch and for energy per phone, we can simply extract those values from a reference audio and overwrite the predictions that the TTS makes. An overview of this process is shown in Figure \ref{fig:overview}. This comes however with two problems: 1) We need accurate alignment of audio to phones on the reference sample and 2) We need to adapt the scale of those values to the new speaker.

To solve the first issue, we train an aligner as an automatic speech recognition system (ASR) using CTC \cite{graves2006connectionist}
and a backtranslation loss inspired by \cite{perez2021vrain}. To get alignments from ASR, we sort the posteriograms and apply monotonic alignment search (MAS) \cite{kim2020glow}. We find that with this lightweight aligner, it is possible to finetune on individual samples in less than a second on a CPU before attempting to clone them. This improves the accuracy of phone boundaries and thus also of the averaged pitch and energy values significantly over just using a neural aligner trained on different data.
And to solve the issue of absolute values for the pitch and energy curves being speaker-dependent, we normalize those values by dividing them by the average of the sequence that they occur in. The resulting curves are no longer speaker dependent. This however introduces the problem of the absolute register of an utterance getting lost. Our solution to this problem lies in the ensemble of speaker embeddings we use, which latently encode the pitch scale. We find that expressive speech and standard speech of the same speaker lead to overlapping, but shifted clusters in the speaker embedding space of the ensemble. So by supplying utterance level speaker embeddings, the normalized representations can automatically be de-normalized for a different speaker.

\subsection{Training Procedure}

Because we intend to train our model on multilingual data for greater diversity in prosody, we represent the inputs to our system as articulatory feature vectors to share knowledge of phones between languages with differing phonemesets, as has been proposed in \cite{lux2022laml}. Furthermore we use the LAML procedure to train the TTS, which has been introduced in the same work. This means that through a sampling strategy in training with multilinguality in mind, we are able to share as much knowledge of the acoustic model between languages as possible.
The aligner is trained on the same data, using a similar procedure to improve it's ability to finetune to unseen data rapidly. This results in a very robust phone-recognizer.

\section{Experimental Setup}
\subsection{Resources}
The main dataset which we use for all of our evaluations is the ADEPT dataset \cite{torresquintero21adept}, which has been specifically designed to evaluate prosody cloning in TTS. 
To train the TTS and the aligner, we use the same datasets. We assume that the greater the diversity in the training data is, the better both the prosody cloning will be able to handle extreme values for pitch, energy and durations and the speaker modelling will be at performing the zero-shot multispeaker task. Hence we use a multilingual model as the basis and train it on a total of 12 languages. For English, we use the  Blizzard Challenge 2011 dataset \cite{king2011blizzard},  LJSpeech  \cite{ljspeech17}, LibriTTS \cite{zen2019libritts}, HiFi-TTS \cite{bakhturina2021hi} and VCTK \cite{veaux2017superseded}. For German we use the HUI-Audio-Corpus-German \cite{puchtler2021huiaudiocorpusgerman} and the Thorsten corpus \cite{muller_thorsten_2021_5525342}. Spanish includes the Blizzard Challenge 2021 dataset \cite{ling2021blizzard} and the CSS10 dataset \cite{css10}, from which we also use the Greek, Finnish, French, Russian, Hungarian and Dutch subsets. The Dutch and French subsets of the Multilingual LibriSpeech \cite{pratap2020mls} are also included, as well as its Polish, Portuguese and Italian subsets. To keep the computational cost manageable, we only use a maximum of 20,000 randomly chosen samples per corpus. This leaves us with a total amount of 400 hours of training data with the highest diversity that we could muster.

\subsection{Experiments}

The ADEPT dataset comes with its own set of benchmark tasks, which we did however not use, since they require vast amounts of human evaluations, for which we lack the capacity. Instead we designed our own experiments in order to verify each of our contributions,  inspired partly by \cite{skerry2018towards}.

In order to measure whether the cloning of the prosody or the cloning of the voice alone or in conjunction affect the quality of the synthetic audio, we use an ASR system to measure the intelligibility by calculating the Phone Error Rate (PER) that the ASR achieves on synthetic speech produced using different configurations and compare it to the PER on human speech. For the ASR we use the architecture described in \cite{karita2019comparative} and the English recipe from the IMS-speech resource \cite{denisov2019ims}.

We use Mel Spectral Distortion (MSD), which is a derivation of Mel Cepstral Distortion \cite{kubichek1993mel} with Dynamic Time Warping (DTW) \cite{berndt1994using} and the F$_0$ Frame Error (FFE) \cite{nakatani2008method} to asses the effectiveness of style cloning when using the same voice as suggested in \cite{skerry2018towards}. The FFE consists of the Gross Pitch Error (GPE) and the Voicing Decision Error (VDE). Our formula for calculating MSD is defined in Equation \ref{eq:mcd}, where $x$ and $y$ are log mel scaled spectrograms, $DTW$ is an algorithm that compares every timestep $t1$ from sequence $x$ to every timestep $t2$ from sequence $y$ to minimize the sum of results of the $dist$ function, which denotes euclidean distance between vectors. $T_x$ denotes the total amount of timesteps in $x$.

\begin{equation}\label{eq:mcd}
\displaystyle
    \text{MSD}(x, y) = \cfrac{DTW(\sum\limits_{t}{dist(x_{t1},y_{t2})})}{T_x}
\end{equation}

We calculate FFE according to Equation \ref{eq:ffe} and it consists of the set of VDEs defined in Equation \ref{eq:vde} and the set of GPEs defined in Equation \ref{eq:gpe}. $x$ and $y$ are pitch contours with unvoiced segments referring to 0 and $T_x$ denotes the total amount of timesteps in $x$. Individual timesteps are denoted as $t$. Intuitively, MSD measures the deviation from a reference speech signal independent of the time-axis, FFE measures how accurately phonation is occurring in a signal and GPE measures the deviation of the absolute pitch.

\begin{equation}\label{eq:ffe}
    \text{FFE}(x, y) =  \cfrac{||\text{VDE}(x,y) \cup \text{GPE}(x,y)||}{T_x}
\end{equation}

\begin{equation}\label{eq:vde}
    \text{VDE}(x, y) = \{t \text{ where } (x_t = 0) \neq (y_t = 0)\}
\end{equation} 

\begin{equation}\label{eq:gpe}
    \text{GPE}(x, y) = \{t \text{ where } \neg (x_t \cdot 0.8 \leq y_t \leq x_t \cdot 1.2)\} 
\end{equation}

To evaluate the perceptive qualities of the cloning procedure in conjunction with the voice cloning, we conduct an online study where raters are asked to indicate how similar a synthetic sample sounds compared to a human reference sample on a scale of 1 to 5. Raters were asked to only take the prosody of the samples into account and ignore the voice of the speaker. The samples for this study were chosen from the Marked Tonicity subset of the ADEPT corpus. All synthetic samples in the study are produced in voices that have not been seen during training and are adapted to in a zero-shot multispeaker fashion. We measure the impact of two different factors by evaluating three types of synthetic samples. For the first category of samples, the voice of the speaker is cloned, but the prosody is left unconditional. For the second category, the voice is cloned and additionally the prosody is cloned. This should give us an indication of how much of a difference the prosody cloning makes perceptually. And the third category clones the prosody, but uses a different voice than the prosody reference. The difference between the second and third category shows whether the speaker embedding and the prosodic parameters are disentangled or whether the normalization of the prosodic parameters followed by the implicit de-normalization using the speaker embedding fails.

While the previous experiment measures the effectiveness of exchanging the prosody regardless of the voice, the final experiment measures the effectiveness of exchanging the voice regardless of the prosody used. For this we measure the speaker similarity across multiple samples spoken by unseen speakers with and without prosody cloning. A difference in absolute speaker similarity that is smaller than the standard deviation would indicate that the prosody cloning has no significant impact on the voice cloning. To measure the speaker similarity objectively, we compute the cosine similarity between speaker embeddings extracted from the reference and the synthetic sample. This is a technique that is frequently applied in zero-shot speaker identification and even speaker verification \cite{wilkinghoff2020open, dehak2010front}.

\section{Results and Analysis}
\subsection{Quantitative Results}
\subsubsection{Degradation of Speech Intelligibility}
The results of our objective evaluation of intelligibility are shown in Table \ref{wer_results}. The voice used for the speaker or whether the prosody is cloned or not only have a marginal impact on the resulting PER. Overall, the synthetic systems perform almost on par with the human speech in terms of intelligibility and the highest pairwise difference between the synthetic types is 0.03 PER, which is negligible. 

\begin{table}[h]
\centering
\begin{tabular}{ l | c } 
\noalign{\hrule height 1pt}
\textbf{Speech Type}                        & \textbf{PER} \\
\noalign{\hrule height 0.5pt}
Human Reference                                      & 12.6\% \\  
TTS - same voice, uncloned prosody   \hspace{.5cm}    & 12.7\% \\
TTS - same voice, cloned prosody              & 12.7\% \\
TTS - different voice, cloned prosody         & 12.7\% \\
\noalign{\hrule height 1pt}
\end{tabular}
\caption{\label{wer_results}
Phone-Error-Rate of an ASR system on the ADEPT corpus for human speech and different synthetic speech configurations, which have not seen the speakers or texts during training.}
\end{table}

\subsubsection{Objective Prosodic Similarity}
Table \ref{pros_results} shows that the cloned style is significantly closer to the reference than a sample that is conditioned on an utterance embedding globally both in terms of MSD and FFE. The reduction in FFE when prosody cloning is applied is almost 3.7 times, which clearly underlines its effectiveness.

\begin{table}[h]
\centering
\begin{tabular}{ l | c | c } 
\noalign{\hrule height 1pt}
\textbf{Speech Type}              & \textbf{MSD} & \textbf{FFE} \\
\noalign{\hrule height 0.5pt}
same voice -  uncloned prosody \hspace{.5cm}          & 25.63       & 48.43\%   \\
same voice -  cloned prosody                               & 5.59        & 13.09\%   \\
\noalign{\hrule height 1pt}
\end{tabular}
\caption{\label{pros_results}
Mel Spectral Distortion and F$_0$ Frame Error for comparing the prosody of a TTS conditioned globally on the speaker and a TTS that additionally clones the prosody to expressive human references. Displayed is the average score of all samples of the ADEPT dataset. The voice of the human speaker was cloned in all cases, since we are dealing with absolute pitch values, which are speaker dependent.}
\end{table}

\subsubsection{Impact on Voice Cloning}
In Table \ref{spk_sims} we show the average cosine similarities of speaker embeddings extracted from human references compared to different configurations of synthetic speech. The speaker embeddings that are being compared are extracted with the same method that we use to get speaker embeddings as conditioning signal, i.e. an ensemble of two state-of-the-art speaker identification models. The average similarity of the synthetic speech with cloned prosody is marginally higher for both the male and the female speaker from the ADEPT dataset. This is possibly the case, because the speech that the scores were calculated from is very expressive and usually has a high pitch range, which is a factor that impacts the speaker embedding. The low standard deviation indicates a concise distribution of values. Compared to intra-human speaker similarity all synthetic speech configurations perform sufficient in the zero-shot multispeaker voice cloning task, regardless of cloning the prosody.

\begin{table}[h]
\centering
\begin{tabular}{ l | l | c | c } 
\noalign{\hrule height 1pt}
\textbf{Speaker}              & \textbf{Cloned} & \textbf{Cosine Similarity} & $\sigma$\\
\noalign{\hrule height 0.5pt}
male & yes   & 0.842 & $\pm0.030$ \\
male & no   & 0.835 & $\pm0.032$ \\
female & yes \hspace{.5cm}   & 0.819 & $\pm0.029$ \\
female & no \hspace{.5cm}   & 0.813 & $\pm0.022$ \\
\noalign{\hrule height 1pt}
\end{tabular}
\caption{\label{spk_sims}
Similarity of all samples from the ADEPT corpus spoken by the two speakers of the ADEPT corpus with and without prosody cloning. Synthetic speech is compared to the human reference, speaker identity of the human reference is always cloned. The human samples in the ADEPT corpus compared to each other reach a similarity of 0.89 for the male speaker and 0.87 for the female speaker, showing the upper bound.}
\end{table}

\subsection{Human Evaluation of Prosody}
Table \ref{study} shows the average similarity rating of 32 participants, who each rated 10 samples per category. So a total of 320 ratings per category are summarized to one score. A slight degradation is expected when the utterance embedding is exchanged, since the range of prosodic values no longer matches the reference. The impact is however very small compared to the massive difference to the similarity that a globally conditioned sample achieves. Furthermore, the standard deviation indicates a major overlap in ratings for the models with cloned prosody, but only a very minor overlap with the model that does not clone the utterance on a phone level.

\begin{table}[h]
\centering
\begin{tabular}{ l | c | c } 
\noalign{\hrule height 1pt}
\textbf{Model}              & \textbf{Similarity} & $\sigma$\\
\noalign{\hrule height 0.5pt}
same voice, cloned prosody   & 4.13 & $\pm0.95$ \\
different voice, cloned prosody   & 3.93 & $\pm1.08$ \\
same voice, uncloned prosody   & 2.09 & $\pm1.09$ \\
\noalign{\hrule height 1pt}
\end{tabular}
\caption{\label{study}
Human rating of similarity between human and synthetic speech on a scale from 1 (dissimilar) to 5 (exact match). Raters were asked to only take the prosody into account and ignore any deviation in voice.}
\end{table}

\subsection{Analysis}
\begin{figure}[h]
    \centering
    \includegraphics[width=.7\linewidth]{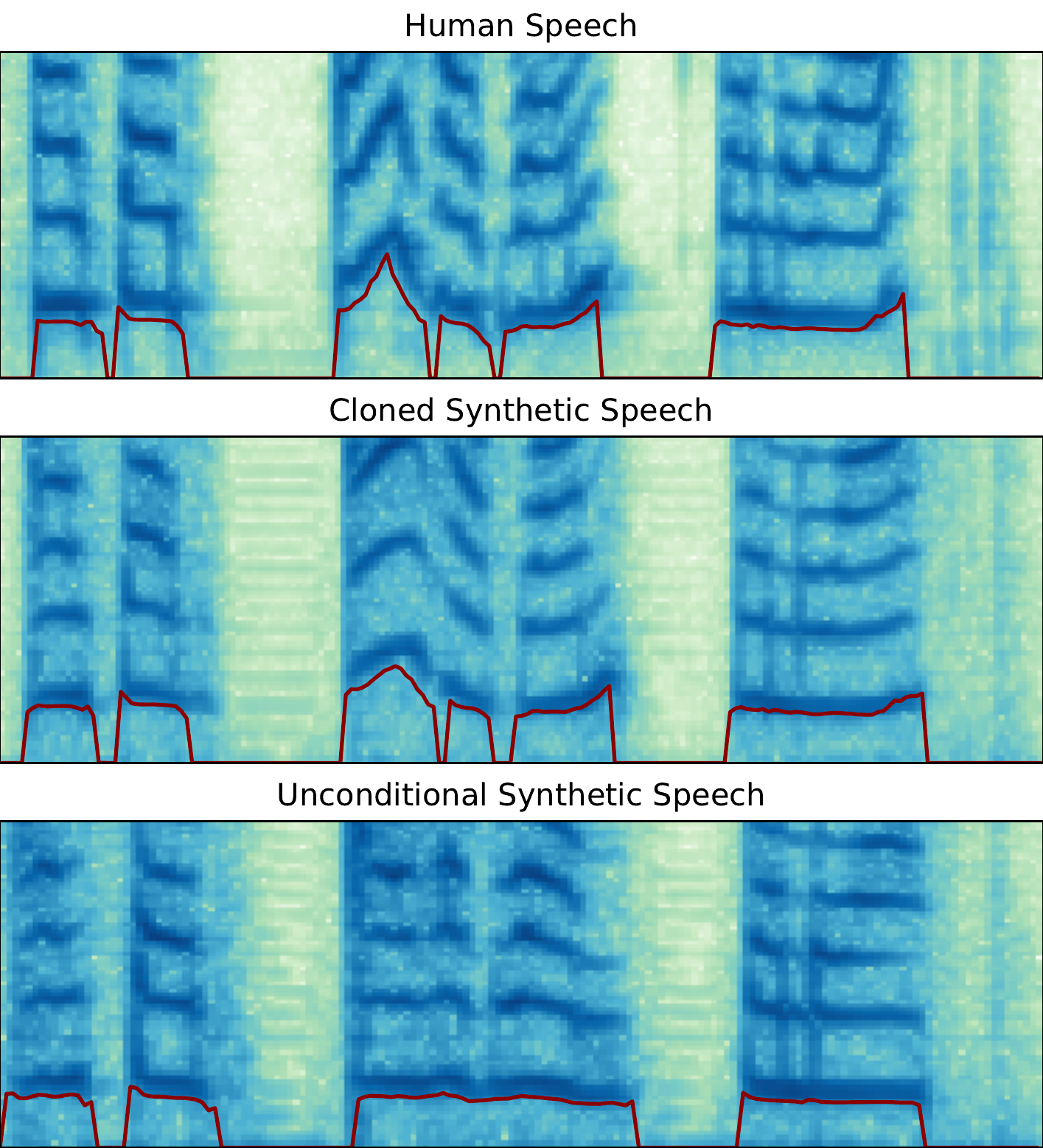}
    \caption{Spectrogram excerpts of the same sentence uttered in an expressive way by a human, the TTS cloning the human and the TTS without cloning. The pitch curve is displayed in red. The voice reference of the synthetic samples is the same as the (unseen) human speaker.}
    \label{fig:example}
\end{figure}

Figure \ref{fig:example} shows an instance of samples which are used for measuring the objective prosodic similarity and the human perception of prosodic similarity. While the pitch contour clearly is not an exact match to the human reference, the general trends are accurately mimicked. Furthermore the absolute value range is accurate except for the peak directly after the first pause, which is sharper in the human speech. So despite the normalization of the pitch curves removing the information about the absolute value range, the utterance level speaker embedding seems to be sufficient to reconstruct it. We refer readers to our demo linked at the beginning of the paper to explore the perceptive performance themselves.

\begin{figure}[h]
    \centering
    \includegraphics[width=.7\linewidth]{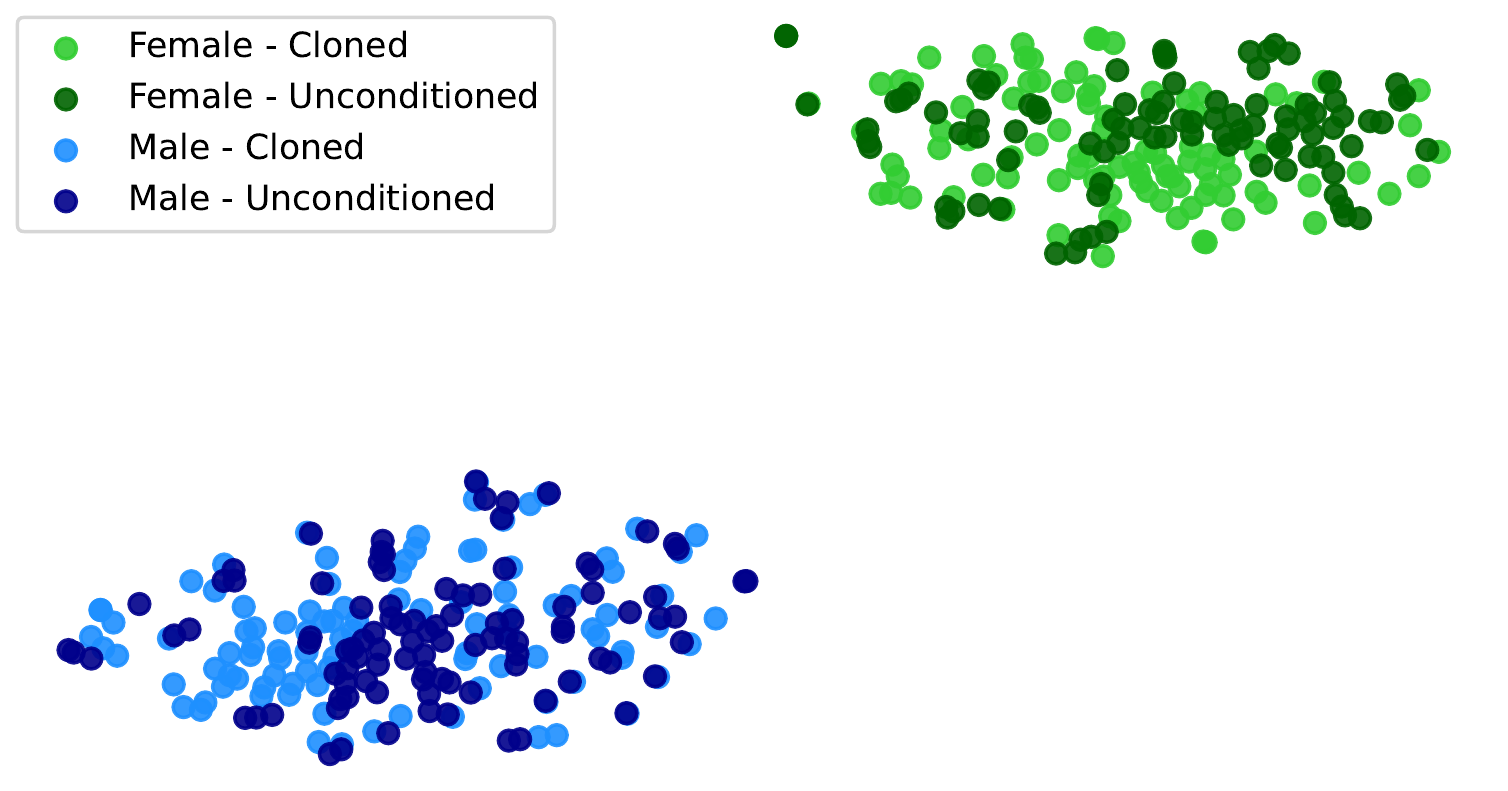}
    \caption{Speaker embedding visualization of the two speakers of the ADEPT dataset with and without cloning. Both cloned and unconditional speech for every sample from the corpus is included, even the samples of the opposite speaker. The presence or absence of the prosody cloning has no impact on the voice of the synthetic samples, which is congruent with our own perception. The speakers are completely unseen.}
    \label{fig:spk}
\end{figure}

To analyze the voice cloning further, we visualize all of the samples in the expressive ADEPT corpus spoken with zero-shot multispeaker voice adaptation by both speakers in the corpus with and without cloning the prosody in Figure \ref{fig:spk}. Since the projection into the 2D space for visualization is done with t-SNE \cite{van2008visualizing} and there are not many distinctive features in the data, the clusters in the visualization look quite fuzzy. However there is a very clear distinction between the two speakers and essentially no distinction between the versions with and without prosody cloning, which is congruent with the results we get in Table \ref{spk_sims}.

\section{Applications} \label{sec:applications}
To better illustrate the merits of using exact prosody cloning in conjunction with zero-shot voice cloning TTS, we will present and discuss a few examples of use cases in this section, which we demonstrate in our demo page linked above.

\subsection{Preserving Meaning in Resynthesis}
The voice of a speaker holds very sensitive information about the identity of the speaker. Using speaker identification techniques, it is possible to find occurrences of the same speaker in different data-sources and map them to a real person, which might have real world implications for that person. Such techniques can be used for modeling users for targeted advertising, for surveillance and geotracking and many more ethically loaded applications. A promising approach for speaker anonymization using TTS with exchanged speaker embeddings has been proposed recently in \cite{anonymization}. This approach however does not conserve the prosody of an utterance that is being anonymized, which can alter the meaning of a sentence, as is illustrated in the section on preserving meaning in resynthesis in our aforementioned demo. Our prosody cloning mechanism can remedy this in all such applications where speech is modified.

\subsection{Voice Unification}
There are plenty of factors that influence the perception of the spoken interpretation of a text, e.g. in poetry. One of those factors is the voice. In order to do meaningful analysis on the impact of certain prosodic patterns on the perception of a listener, the voice needs to be accounted for and ideally taken out of the equation through debiasing techniques. On the data-side of the problem however, there is usually only one realization of a poem per speaker. Using our suggested method of combining voice-cloning with prosody cloning, we can unify multiple realizations of e.g. the same poem read by different speakers to have the same voice. 
This is illustrated for the case of German poetry in the section on voice unification in our demo. This further demonstrates how our approach is mostly language agnostic.

\subsection{Customization of Voice-Acting}
Hiring professional speakers to read e.g. an audiobook is quite costly. Having multiple professional speakers voice-act different characters in an audioplay is thus usually even more costly. With our proposed approach and the aligner model we propose along with it, it is possible to take an audiobook, align every sentence to the corresponding text, parse from the text which speaker is currently speaking and then re-synthesizing each sentence with a newly chosen voice while retaining the prosody of the professional speaker. We demonstrate this full pipeline in our demo in the section on transforming an audiobook into a customizable audioplay. All of the steps are performed without human intervention. For simplified parsing of who speaks when, we choose a theatre play, where such annotations are quite clear. With this demo application, we made the observation that boundaries between per-speaker-segments in long audios of multiple minutes of speech can sometimes be problematic because of compounding errors along the time-axis if the MAS finds an incorrect path through the posteriors. This can be resolved quite easily however by using an ensemble of multiple predictions made by the same aligner model and averaging the predicted utterance boundaries.

\section{Ethical Considerations}
While technology that allows the cloning of voices and speaking styles has been around for a while \cite{jia2018transfer, wang2018style, skerry2018towards}, the method of extracting and overwriting conditioning signals in a TTS that we propose allows for more nuanced control over the cloned speech, disentangling the speaker from the prosody. While this holds the potential to be misused to impersonate others with more control over how the impersonation sounds, we believe that the applications with a positive impact outweigh this fact. For example the masking of a voice to keep the speaker private, but retain their exact prosody can alleviate many other ethical problems that are introduced by the application of other speech analysis techniques. Achieving voice masking with unchanged prosody is still an unsolved trade-off in current voice sanitization methods \cite{tomashenko2022voiceprivacy}. Furthermore, we plan on experimenting with inaudible audio-watermarking that is encoded into the weights of a TTS as well as deepfake detection models to ensure that synthetic samples can always be easily identified as such.

\section{Conclusions}
We show an approach for cloning the prosody of a reference audio in terms of duration, pitch and energy on a phone level while being able to vary the voice of the TTS in a zero-shot manner. Objective evaluation shows that 1) there is no degradation in quality, 2) the performance of voice cloning is not impacted by prosody cloning and 3) the prosody cloning per phone achieves greatly increased similarity to an expressive human reference than a globally conditioned sample. Human evaluation further shows that the quality of the prosody cloning is high and not impacted by the voice cloning.


\bibliographystyle{IEEEbib}
\bibliography{main}

\end{document}